\documentclass[onecolumn,preprintnumbers,amsmath,amssymb]{revtex4}

\usepackage{graphicx}% Include figure files
\usepackage{subfigure}% Include figure files
\usepackage{dcolumn}% Align table columns on decimal point
\usepackage{bm}% bold math
\usepackage[usenames,dvipsnames]{xcolor}
\usepackage[colorlinks=true,citecolor=Cerulean,linkcolor=RubineRed,urlcolor=Cerulean]{hyperref}

\begin{document}

\title{Shortcuts to adiabaticity in  Fermi gases}% 
\author{Pengpeng Diao$^1$, Shujin Deng$^{1}$, Fang Li$^1$, Shi Yu$^1$, \\
 Aur\'elia Chenu$^2$, Adolfo del Campo$^3$,  and Haibin Wu$^{1,4}$}

\affiliation{$^1$State Key Laboratory of Precision Spectroscopy, East China Normal University, Shanghai 200062, P. R. China\\
$^2$Theoretical Division, Los Alamos National Laboratory, Los Alamos, NM 87544, USA\\
$^3$Department of Physics, University of Massachusetts, Boston, MA 02125, USA\\
$^4$Collaborative Innovation Center of Extreme Optics, Shanxi University, Taiyuan 030006,China
}

\date{\today}

%\affiliation{%
%}%

%\author{}
 %\homepage{}
%\affiliation{
% with \\
%}%
%

\date{\today}% It is always \today, today,
             %  but any date may be explicitly specified

%\pacs{Valid PACS appear here}% PACS, the Physics and Astronomy

\begin{abstract}
Shortcuts to adiabaticity (STA) provide an alternative to adiabatic protocols to guide the dynamics of the system of interest without the requirement of slow driving.
We report the controlled speedup via STA of the nonadiabatic dynamics of  a Fermi gas, both in the non-interacting and strongly coupled, unitary regimes. Friction-free superadiabatic expansion strokes, with no residual excitations in the final state, are  demonstrated in the unitary regime by engineering the modulation of the frequencies and aspect ratio of the harmonic trap.  STA are also analyzed and implemented  in the high-temperature regime, where the shear viscosity plays a pivotal role and the Fermi gas is described by viscous hydrodynamics. \end{abstract}

\maketitle

\section{Introduction}
Developing the ability to tailor the dynamics of complex quantum systems has been a long-time goal across a variety of fields. In addition, this goal is widely recognized as a necessity for the advancement of quantum technologies.  However, the presence of strong correlations between constituent particles hinders the understanding and control of the time evolution of many-body systems. In view of this complexity barrier, emergent symmetries can play a pivotal role to simplify the dynamics far away from equilibrium and its control.

A paradigmatic test-bed of nonequilibrium many-body physics in the laboratory is provided by ultracold Fermi gases. Interatomic  interactions  in these systems can be considered of  zero range. Using the Feshbach resonance technique ~\cite{Chin10}, the strength of the interactions can be varied from zero value, creating an ideal Fermi gas, to a divergent interaction, leading to  the unitary regime where the scattering length is infinite.
Incidentally, these two extreme regimes are characterized by scale invariance as an emergent symmetry, which is broken for any finite value of the interaction strength.
The appearance of scale invariance is crucial to describe the strongly-coupled unitary Fermi gas and leads to universality in the thermodynamics and hydrodynamics of the system ~\cite{Ku12,Kinast05,Ho04,Makotyn14,Cao11}. Moreover, as a dynamical symmetry, scale invariance  relates the time evolution of the unitary Fermi gas  to equilibrium properties of the system. For instance, the evolution of local correlation functions such as the density profile of the system becomes self-similar. As a result, it  can be simply described by a scaling of the coordinates with a time-dependent scaling factor. The connection between properties in- and out-of-equilibrium greatly reduces the complexity of the time evolution and has spurred developments in understanding intricate few-body and many-body dynamics.
Beyond the study of Fermi gases,
scale invariance has proved extremely useful in the exploration of ultracold atomic gases in time-dependent harmonic traps and provides the means to analyze time-of-flight measurements \cite{Castin96,Kagan96}. We can thus expect that it can be harnessed to provide fast control of quantum systems far-away from equilibrium.

Shortcuts to adiabaticity (STA)  aim at speeding up the  evolution of a system in a controlled way without the requirement of slow driving~\cite{Chen2010b,Torrontegui2013}.
As a general control tool, STA have found broad applications across a variety of fields, such as population transfer \cite{Demirplak2003,Demirplak2005,Demirplak2008,Berry2009,Chen2010b,Ruschhaupt2012,Zhang13,Masuda2015,Du2016}, quantum thermodynamics \cite{Deng13,Campo2014a,Beau16,Funo17,Deng18Sci},  the control of critical systems \cite{Campo2012a,Takahashi2013a,Damski2014,Saberi2014,Campo2015,Takahashi2017}, and fast and robust quantum transport \cite{Masuda2009,Torrontegui2011,An2016}.
Several techniques have been developed for the design of STA. Counterdiabatic driving \cite{Demirplak2003,Demirplak2005,Demirplak2008,Berry2009} constitutes a universal approach provided that the spectral properties of the system are known. When this is not the case, alternative methods are desirable. Prominent examples, with complementary advantages and varying range of applicability, include the fast-forward technique \cite{Masuda2009,Masuda2011,Masuda2014}, the use of invariant of motions and scaling laws \cite{Chen2010b,delcampo11epl,Choi2011b,Jarzynski13,delcampo2013,Choi2013,Deffner2014,Jarzynski2017},  classical flow fields \cite{Patra2017},  the existence of Lax pairs in integrable systems \cite{Okuyama2016}, and counterdiabatic Born-Oppenheimer dynamics \cite{Callum18}.

Progress to control trapped ultracold gases and many-body quantum fluids has been facilitated by the use of dynamical symmetries and the associated scaling laws \cite{Chen2010b,Muga09,delcampo11epl,Campo2011,Choi2011,Campo2012b,delcampo2013,Deffner2014}.
In this context, STA were first demonstrated in the laboratory with a thermal atomic cloud \cite{Schaff2010}, and soon after using a Bose-Einstein condensate,  well described by mean field theory \cite{Schaff2011a,Schaff2011}. Theoretical work indicated that STA could be applied to arbitrary quantum fluids with scale invariant symmetry \cite{Campo2011,Campo2012b,delcampo2013,Deffner2014} and STA were later implemented to control an effectively one-dimensional atomic cloud with phase fluctuations \cite{Rohringer2015}.
Recently, we have demonstrated that STA can as well be applied in the strongly-coupled regime, using a three-dimensional (3D) anisotropic Fermi gas at unitarity as a test-bed~\cite{Deng18pra}. The superadiabatic quantum friction suppression in finite-time thermodynamics has further been demonstrated in this system~\cite{Deng18Sci}.

In this article, we present a detailed study of STA for the driving of Fermi gases  both in the noninteracting regime and at unitarity. In particular, we show that it is possible to implement STA by engineering exclusively the time-dependent anisotropic trap, this is, without additional auxiliary controls. Further, we  explore the superadiabatic control of a  unitary Fermi gas in the high temperature regime.  At finite-temperature, the shear viscosity cannot be neglected~\cite{Cao11,Elliott14}, as it substantially affects the nonadiabatic dynamics of the system.  The evolution can then be described by viscous hydrodynamics  and the well known ``elliptic" flow at unitarity ~\cite{OHara02} will be changed. While the effect of viscosity can limit the performance of STA in  anisotropic expansions and compressions, it vanishes whenever the dynamics is isotropic. Our work shows that STA can be broadly applied in ultracold atomic gases across different interaction regimes and in the presence of viscosity.

The paper is organized  as follows. In Section \ref{sec1} we characterize scale invariance as  dynamical symmetry  governing the dynamics of a Fermi gas  at low temperature both in the noninteracting and unitary regimes. In Section \ref{sec2} we present the experimental demonstration of STA for the expansion and compression of a Fermi gas in  the strongly interacting regime.  The dynamics taking shear viscosity into consideration at high temperature is studied in Section \ref{sec2a}.  We  conclude with a  summary and outlook in  Section \ref{sec3}.

\section{Design of shortcuts to adiabaticity in ultracold Fermi gases} \label{sec1}

 The noninteracting and unitary Fermi gases are both scale invariant, but with different scaling equations governing their dynamics.
 In the noninteracting case, the equations governing the evolution  along different axes are decoupled  due to the lack of collisions.  By contrast,  the dynamics along different axes are strongly coupled at unitarity.

\subsection{Noninteracting Fermi gas}
Consider a 3D noninteracting Fermi gas confined in a time-dependent anisotropic harmonic trap, described by the Hamiltonian
\begin{equation}
\hat{H}(t)=\sum_{i=1}^{N}\bigg[-\frac{\hbar^2}{2m}\nabla_i^2+\frac{1}{2}m\bigg(\omega_x^2(t)\, \hat{x}_i^2+\omega_y^2(t)\, \hat{y}_i^2+\omega_z^2(t)\, \hat{z}_i^2\bigg)\bigg].~\label{eq1 }
\end{equation}
We focus on the evolution of the system following a time-modulation of the trap frequencies $\omega_j$ ($j=x,y,z$) to induce an expansion or compression of the gas.
The system exhibits scale invariance and the dynamics in this  regime can be described by time-dependent scaling factors $b_j(t)$ ($j=x,y,z$) given by
\begin{equation}
b_{x}(t)=\left[\frac{\left\langle \hat{R}_x^2(t)\right\rangle}{\left\langle \hat{R}_x^2(0)\right\rangle}\right]^{\frac{1}{2}},
\quad b_{y}(t)=\left[\frac{\left\langle \hat{R}_y^2(t)\right\rangle}{\left\langle \hat{R}_y^2(0)\right\rangle}\right]^{\frac{1}{2}},
\quad b_{z}(t)=\left[\frac{\left\langle \hat{R}_z^2(t)\right\rangle}{\left\langle \hat{R}_z^2(0)\right\rangle}\right]^{\frac{1}{2}}.
\end{equation}
Thus, the scaling factors are defined in terms of the variance of the collective coordinates
\begin{equation}
\hat{R}_x^2=\sum_{i=1}^N \hat{x}_i^2, \quad \hat{R}_y^2=\sum_{i=1}^N \hat{y}_i^2, \quad \hat{R}_z^2=\sum_{i=1}^N \hat{z}_i^2 
\end{equation}
measured in the state of the cloud, and describe the evolution of the density profile of the trapped atomic cloud that the ideal Fermi gas forms.
Their dynamics is dictated by the uncoupled equations, for each Cartesian coordinate,
\begin{equation}
\ddot{b}_j+\omega_j^2(t)b_j=\frac{\omega_{j,0}^2}{b_j^{3}},~~~(j=x,y,z)~\label{eq2},
\end{equation}
with  boundary conditions~$b_j(0)=1$ and $\dot{b}_j(0)=0$. As a result, the evolution of the cloud size is completely determined by the time-dependent trapping frequencies.

A simplified scenario concerns the expansion from an isotropic trap, where a single scaling factor $b(t)$ suffices to completely describe the evolution of the system.
For the cloud to follow a given desirable trajectory described by $b(t)$, the trap frequencies are to be modulated as  \cite{Campo2011,delcampo2013}
\begin{equation}
\omega_j^2(t)=\frac{\omega_{j,0}^2}{b^{4}}-\frac{\ddot{b}}{b}.\,~\label{eq3}
\end{equation}
The existence of scaling laws thus makes possible to control the dynamics of the system via STA, speeding up the adiabatic transfer between two many-body stationary states by controlling the aspect ratio of the frequencies~\cite{Campo2011,delcampo2013,Deng18pra}.

An important application of STA is the engineering of thermodynamic processes to extract the maximum available work in the minimum possible time \cite{Deng13,Campo2014a,Beau16,Funo17}. In a unitary process, the mean work equals the  change in energy between the final and initial state \cite{Talkner07}. To optimize a process using STA, it suffices to characterize the nonadiabatic mean-energy. For a non-interacting 3D Fermi gas, the different degrees of freedom decouple. The total energy is thus the sum of the individual energy along each  degree of freedom.   For the initial state $\langle H(0)\rangle=3m\omega_0^2\sigma_0^2$, where $\sigma_0$ is the mean square cloud size, the adiabatic limit of (\ref{eq3}) is reached when $\dot\omega(t)/\omega(t)^2\ll1$~\cite{LR69}. Then, for a state initially at thermal equilibrium in an isotropic trap with frequency $\omega_{0}$, the adiabatic scaling factor is given by $b_{ad}(t)=\sqrt{\omega_{0}/\omega(t)}$. The nonadiabatic evolution of the mean energy $\langle H(t)\rangle$ and mean work $\langle W(t)\rangle$ read
\begin{eqnarray}
\langle H(t)\rangle&=&\frac{Q^*(t)}{b_{ad}^2}\langle H(0)\rangle=Q^*(t)3m\omega_0\omega(t)\sigma_0^2,\\
\langle W(t)\rangle&=&\langle H(t)\rangle-\langle H(0)\rangle=\left(Q^*(t)\frac{\omega(t)}{\omega_0}-1\right)\langle H(0)\rangle,
\end{eqnarray}
where~ $Q^*(t)$~ is the nonadiabatic factor given by \cite{Jaramillo16,Beau16}
\begin{equation}
Q^*(t)=b_{ad}^2\bigg[\frac{1}{2b^2}+\frac{\omega(t)^2}{2\omega_0^2}b^2+\frac{\dot{b}^2}{2\omega_0^2}\bigg].
\end{equation}

Note that in the adiabatic limit, $\ddot{b}\approx0$, the scaling factor $b(t)$ approaches its adiabatic value $b_{ad}(t)=\sqrt{\omega_0/\omega(t)}$ and the nonadiabatic factor $Q^*(t)$ equals unity. In this case, the mean energy  is set by its adiabatic value $\langle H(t)\rangle=\langle H(0)\rangle \omega(t)/\omega_0$ and no quantum friction exists. Values of $Q^*(t)> 1$ indicate deviations from adiabatic dynamics and can be associated with quantum friction~\cite{Feldmann06}, which vanishes whenever $Q^*(t)=1$.

\subsection{Unitary Fermi gas}\label{sec12}
The unitary Fermi gas is reached in the strongly-interacting regime, where  the divergent scattering length at resonance leads to different dynamics from the noninteracting Fermi gas. 
A 3D anisotropic unitary Fermi gas in a time-dependent anisotropic harmonic trap is described by the Hamiltonian
\begin{equation}
\hat{H}(t)=\sum_{i=1}^{N}\bigg[-\frac{\hbar^2}{2m}\nabla_i^2+\frac{1}{2}m\bigg(\omega_x^2(t)\,  \hat{x}_i^2+\omega_y^2(t)\,  \hat{y}_i^2+\omega_z^2(t) \,\hat{z}_i^2\bigg)\bigg]
+\sum_{i<j}U(\mathbf{r}_i-\mathbf{r}_j),
\end{equation}
where $U(\mathbf{r}_i-\mathbf{r}_j)$ describes zero-range pairwise interactions with a divergent scattering length.
In particular, $U(\mathbf{r}_i-\mathbf{r}_j)$ is a homogeneous function with the same scaling dimension as the kinetic energy operator.
In contrast to the noninteracting Fermi gas, the dynamics along different axes for the strongly interacting Fermi gas at resonance is strongly coupled. The evolution of the cloud size at unitarity is governed by
\begin{equation}
\ddot{b}_j+\omega_j^2(t)b_j=\frac{\omega_{j,0}^2}{b_j\Gamma^{2/3}},
\label{eq7}
\end{equation}
where $b_{j}(t),(j=x,y,z)$ are the scaling factors corresponding to this regime and $\Gamma(t)=b_x(t)b_y(t)b_z(t)$ is the scaling volume factor.

Our approach to realize the superadiabatic control is based on the counterdiabatic driving technique~\cite{delcampo2013,Beau16}, which relies on first designing a desirable reference adiabatic evolution and subsequently identifying the consistent conditions  to describe its exact nonadiabatic quantum dynamics, in a predetermined time $\tau$.

To design the reference evolution of the cloud, let $\{\omega_{j,0}|j=x,y,z\}$ denote the frequencies of the anisotropic harmonic trap at $t=0$. Similarly, let $\{b_{j,\tau}|j=x,y,z\}$ denote the target scaling factors upon completion of an expansion or compression
stroke of duration $\tau$. The required boundary conditions are as follows
\begin{eqnarray}
\omega_j(0)&=&\omega_{j,0}, ~~~\omega_j(\tau)=\omega_{j,0} /b_{j}(\tau)^2,\nonumber\\
\dot{\omega}_{j}(0)&=&0,~~~~~~\dot{\omega}_{j}(\tau)=0,\nonumber\\
\ddot{\omega}_{j}(0)&=&0,~~~~~~\ddot{\omega}_{j}(\tau)=0. \nonumber
\end{eqnarray}
Satisfying these boundary conditions, we choose the time-dependent trap frequencies via the polynomial Ansatz $\omega_j(t)=\omega_{j,0}\sum c_n(\frac{t}{\tau})^n$,
\begin{equation}
\omega_j(t)=\omega_{j,0}+[\omega_j(\tau)-\omega_{j,0}]\bigg[10\left(\frac{t}{\tau}\right)^3-15\left(\frac{t}{\tau}\right)^4+6\left(\frac{t}{\tau}\right)^5\bigg].
\end{equation}
Using the   adiabatic equations of motion, we determine the reference expansion factor as
\begin{align}
b_j(t)=\frac{\omega_{j,0}}{\omega_j(t)\Gamma^{1/3}(t)}=\frac{\omega_{j,0}}{\omega_{j}(t)}\left[\frac{\nu(t)}{\nu(0)}\right]^{1/2},
\end{align}
where $\nu(t)=[\omega_x(t)\omega_y(t)\omega_z(t)]^{1/3}$ is the geometric mean frequency.

The above equations describe the evolution in the adiabatic limit under slow driving. Nonetheless, they can represent exact nonadiabatic dynamics under a modified driving protocol by a different time-dependence of the trapping frequencies, i.e., replacing $\omega_j(t)\rightarrow\Omega_j(t)$ where the explicit form of $\Omega_j(t)$ is to be determined. This approach has been studied for the single-particle time-dependent harmonic oscillator and many-body quantum systems. It is generally referred to as local counterdiabatic driving (LCD)~\cite{Beau16}. The requiring driving frequencies are given by
\begin{align}
\Omega_j^2(t)&=\frac{\omega_{j,0}^2}{b_j^2\Gamma^{2/3}}-\frac{\ddot{b}_j}{b_j},\nonumber\\
&=\omega_j^2(t)-\frac{\ddot{b}_j}{b_j}.
\label{LCDfreqanis}
\end{align}
This yields the explicit expression for $\Omega_j^2(t)$  as \cite{Deng18Sci}
\begin{align}
\Omega_j^2(t)&=\omega_j^2-2\left(\frac{\dot{\omega}_j}{\omega_j}\right)^2+\frac{\ddot{\omega}_j}{\omega_j}-\frac{4}{9}\left(\frac{\dot\Gamma}{\Gamma}\right)^2
+3\frac{\ddot{\Gamma}}{\Gamma}-\frac{2}{3}\frac{\dot{\omega}_j\dot{\Gamma}}{\omega_j\Gamma},\nonumber\\
&=\omega_j^2-2\left(\frac{\dot{\omega}_j}{\omega_j}\right)^2+\frac{\ddot{\omega}_j}{\omega_j}+\frac{1}{4}\left(\frac{\dot\nu}{\nu}\right)^2
-\frac{1}{2}\frac{\ddot{\nu}}{\nu}+\frac{\dot{\omega}_j\dot{\nu}}{\omega_j\nu},
\end{align}
which includes the counterdiabatic corrections arising from the time-dependence of $\omega_j$, the geometric mean $\nu$ and their coupling.

According to Ref.~\cite{Deng18Sci}, the nonadiabatic factor and mean work read
\begin{align}
Q^{\ast}(t)&=\Gamma_{ad}^{2/3}\bigg[\frac{1}{2\Gamma^{2/3}}+\frac{1}{6}\sum_{j=x,y,z}\frac{\dot{b}_j^2+\omega_j^2(t)b_j^2}{\omega_{j,0}^2}\bigg],\\
\langle W(t)\rangle&=\langle H(t)\rangle-\langle H(0)\rangle=\bigg[\frac{1}{2\Gamma^{2/3}}+\frac{1}{6}\sum_{j=x,y,z}\frac{\dot{b}_j^2+\omega_j^2(t)b_j^2}{\omega_{j,0}^2}-1\bigg]\langle H(0)\rangle.
\end{align}
The last equation follows from the fact that, for isolated quantum systems evolving under unitary dynamics, the (mean) work reduces to the difference in energy between the final and the initial state \cite{Talkner07}.
 For the special case in which  the time evolution is isotropic, the scaling factors  are set by $b_j(t)=(\omega_{j,0}/\omega_j(t))^{1/2}=b(t)$,  the volume scaling factor simplifies to $\Gamma(t)=b^3(t)=(\omega_{j,0}/\omega_j(t))^{3/2}$  and
\begin{align}
\Omega_j^2(t)=\omega_j^2(t)-\frac{3}{4}\bigg[\frac{\dot{\omega}_j(t)}{\omega_j(t)}\bigg]^2+\frac{1}{2}\frac{\ddot{\omega}_j(t)}{\omega_j(t)}.\label{eq5}
\end{align}
The nonadiabatic factor $Q^{\ast}(t)$ and mean work $\langle W(t)\rangle$ are then given by
\begin{eqnarray}
Q^{\ast}(t)&=&1+\frac{1}{12}\sum_{j=x,y,z}\left(\frac{\ddot{\omega}_j}{\omega_j^3}-\frac{\dot{\omega}_j^2}{\omega_j^4}\right),\\
\langle W(t)\rangle&=&\bigg(\frac{Q^{\ast}(t)}{b_{ad}^2}-1\bigg)\langle H(0)\rangle.
\end{eqnarray}

\section{Experimental implementation of Shortcuts to adiabaticity in ultracold Fermi gases} \label{sec2}
\begin{figure}[htb]
\centering
\includegraphics[width=12cm]{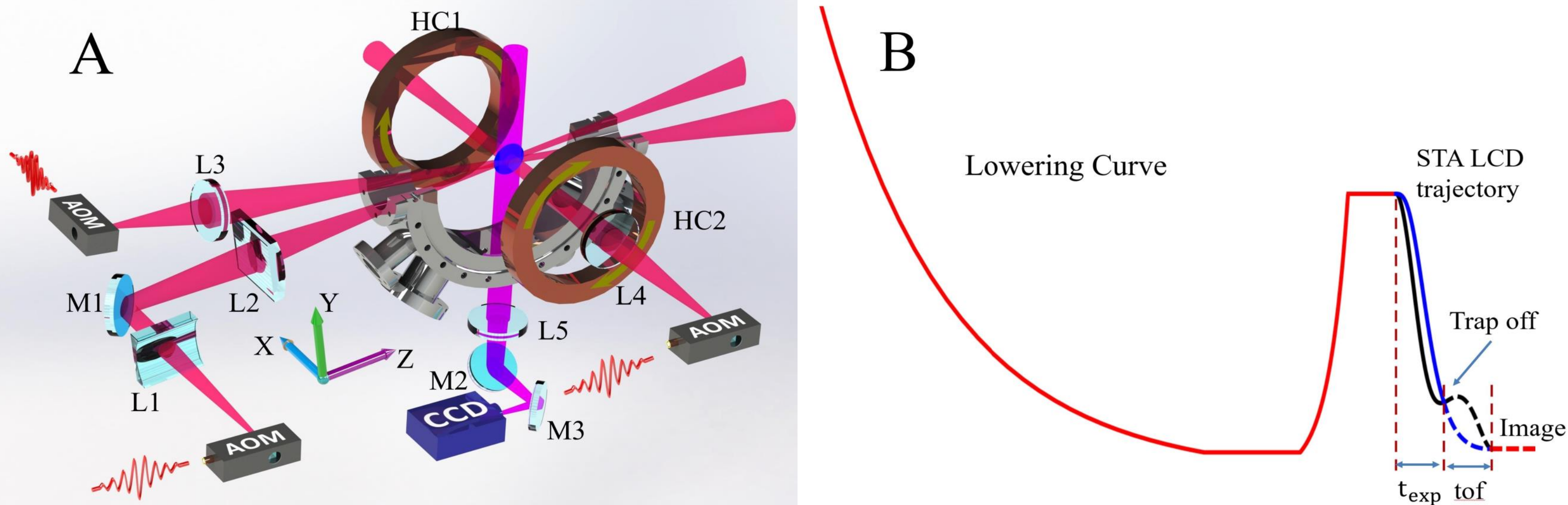}
\caption{{\bf Experimental setup.} (A) Schematic representation of the experimental setup and (B) time-dependence of the beam power during the experimental protocol.  In (B) the red line indicates the beam power during the evaporative cooling, while the blue and black lines represent the power intensity of the two separated beams along a STA expansion stroke. A specially designed optical crossed-dipole trap is formed by two orthogonal far-off resonance laser beams, providing a highly controllable trap frequency. M1-M4, Mirrors; L1-L2, cylindrical lenses; HC1-HC2, Feshbach coils; L3-L4, achromatic lenses; AOM, acousto-optic modulator; tof, time-of-flight. }\label{setup}
\end{figure}

Our experiment is implemented in a 3D anisotropically-trapped unitary quantum gas, with a balanced mixture of $^6$Li fermions in the lowest two hyperfine states  $|\!\!\uparrow\rangle\equiv|F=1/2, M_F=-1/2\rangle$ and $|\!\!\downarrow\rangle\equiv|F=1/2, M_F= 1/2\rangle$. We probe the nonadiabatic expansion dynamics by varying in time  the  harmonic trap frequency. The experimental setup is shown in Fig.~\ref{setup}, and is similar to that in  Ref.~\cite{Deng18Sci}. The atoms are first loaded into an optical dipole trap formed by a single beam. A forced evaporation is performed to cool atoms to quantum degeneracy in an external magnetic field at 832~G. Then, the atoms are transferred to another dipole trap, which consists of an elliptic beam generated by a cylindrical lens along the $z$-axis and a nearly-ideal Gaussian beam along the $x$-axis. The resulting potential has a cylindrical symmetry around $x$. This trap facilitates the accurate tuning of the trap frequencies to control  the anisotropy and  geometry of the atomic cloud. A Feshbach resonance is used to tune the interaction of the atoms either to the non-interacting regime with the magnetic field $B = 528$ G or to the unitary limit with $B = 832$ G.
\begin{figure}[htb]
\centering
\subfigure{ \label{isotropicA}
\includegraphics[width=5cm]{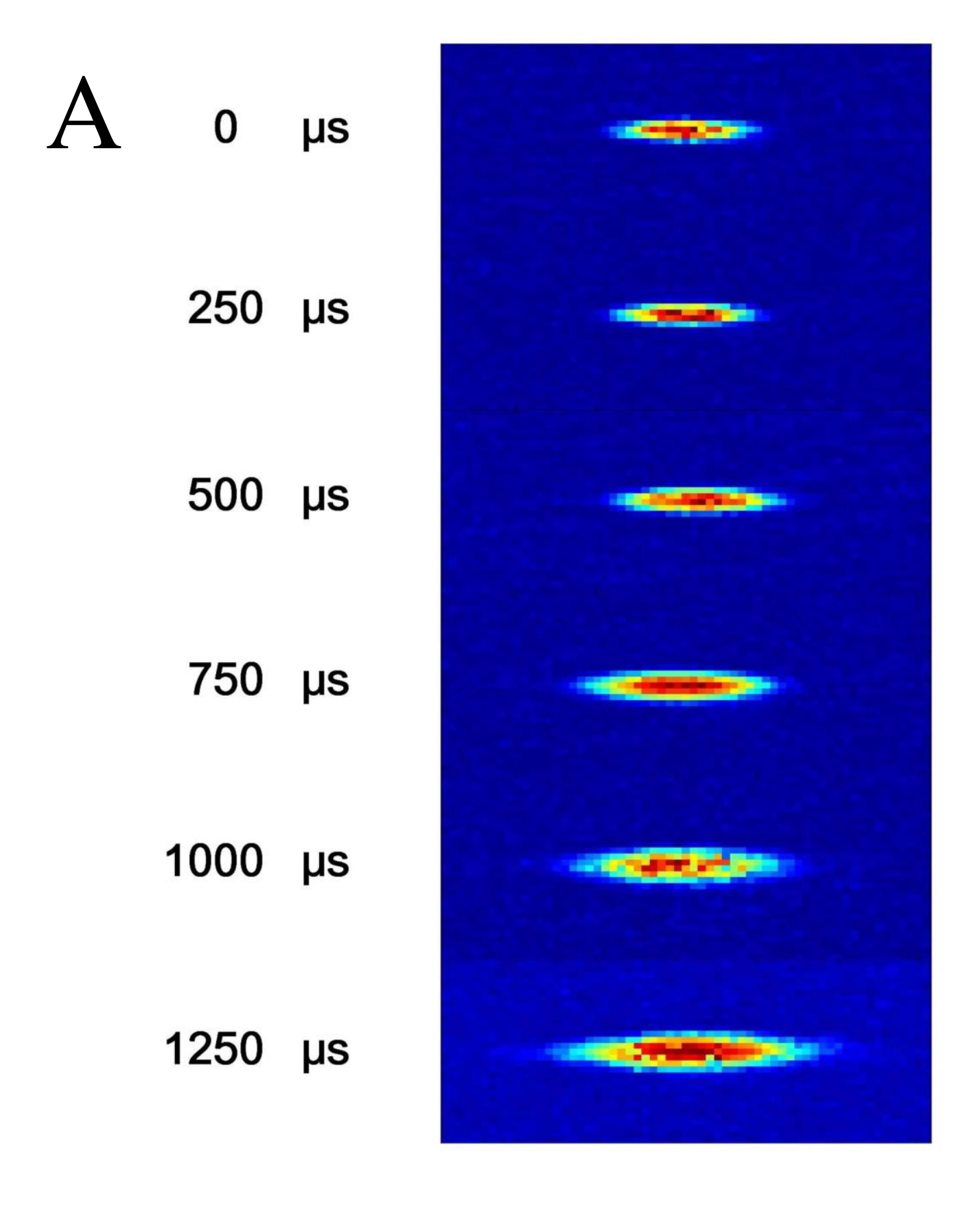}
}
\hspace{0.5cm}
\subfigure{ \label{isotropicB}
\includegraphics[width=8cm]{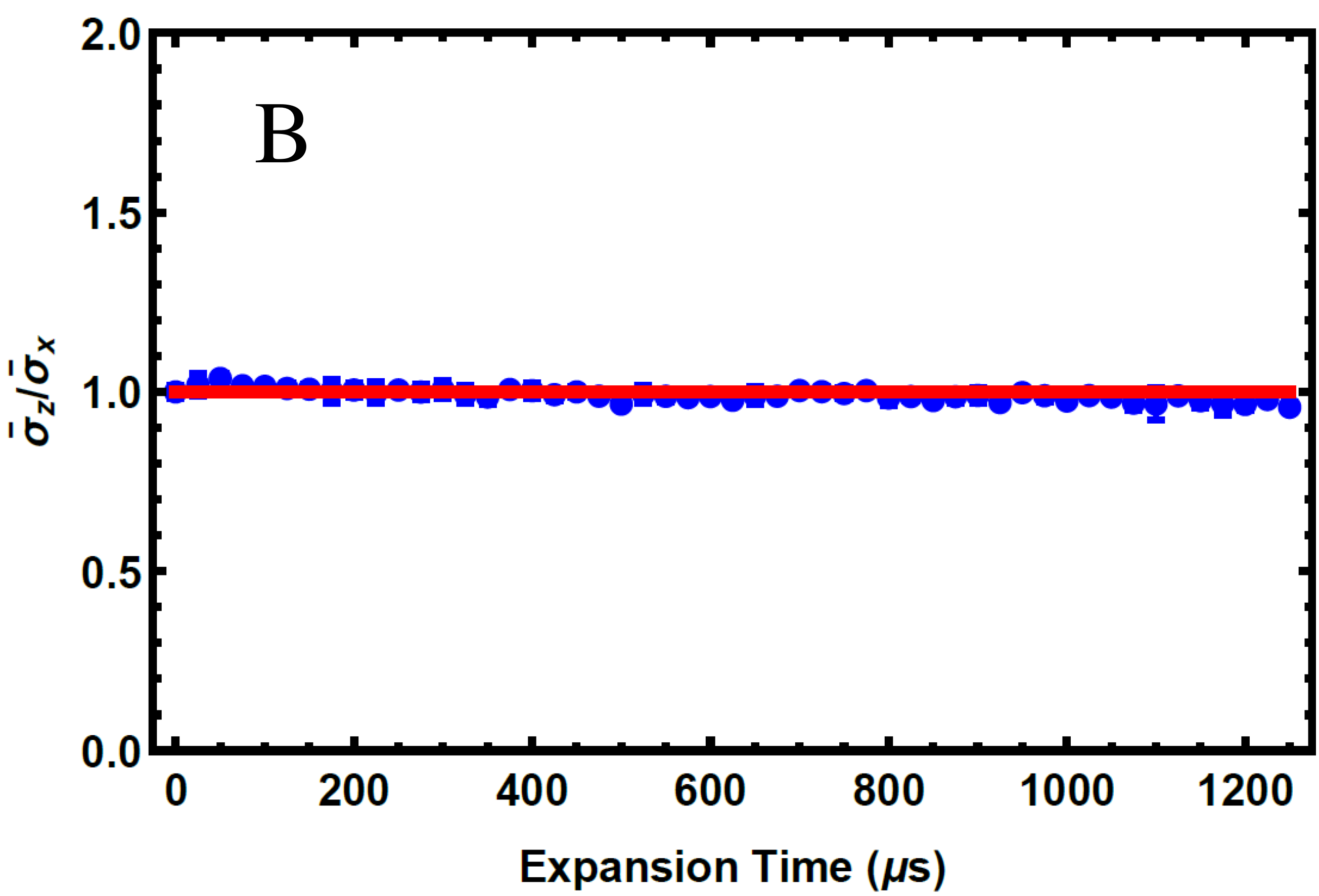}
}
\caption{{\bf STA for an isotropic expansion.} (A) Sequence of density profile images  and (B) aspect ratio of the atomic cloud during a STA. Red dots are the measured results for LCD while the red line denotes a constant value of unity. Error bars represent the standard deviation extracted from the measurement statistics. Measurement data shows that  $\bar\sigma_z$ and $\bar\sigma_x$ are matched during the evolution, showing that the expansion is  isotropic.}\label{isotropic}
\end{figure}
The system is initially prepared in a stationary state of a normal fluid, with $\omega_x(0)=2\pi\times825$ Hz and $\omega_y(0)=\omega_z(0)=2\pi\times230$ Hz. The initial energy of the Fermi gas at unitarity is $E=0.75 (0.1)\,E_F$, corresponding to a temperature $T=0.23(0.02)\,T_F$, where $E_F$ and $T_F$ are the Fermi energy and temperature of an ideal Fermi gas, respectively. Here we focus on the hydrodynamic expansion of a unitary Fermi gas, when   the magnetic field $B = 832$ G. To engineer  an isotropic expansion in an anisotropic trap, the frequency aspect ratio needs to be controlled in the experiment (set up here at 3.59 at the beginning of an STA process). The target final value of the scaling factor $b(\tau)$ is chosen to be 1.5 in a transferring time $\tau=1250\,\mu$s. Snapshots of the density profile of the atomic cloud and its aspect ratio during the expansion are shown in Fig. \ref{isotropic}. We engineer an isotropic expansion via LCD and define two time dependent dimensionless cloud sizes, $\bar\sigma_z=\sigma_z(t)/\sigma_z(0)$ and $\bar\sigma_x=\sigma_x(t)/\sigma_x(0)$, to characterize the time evolution. It is clear that if the expansion is isotropic,  $\bar\sigma_z/\bar\sigma_x$ should be equal to unity at all times.  The measured data of the aspect ratio of the atomic cloud,  presented in Fig. \ref{isotropicB}, confirms that the expansion is isotropic in spite of the anisotropy of the trap.

\begin{figure}[t]
\centering
\subfigure{ \label{fig1B}
\includegraphics[width=8cm]{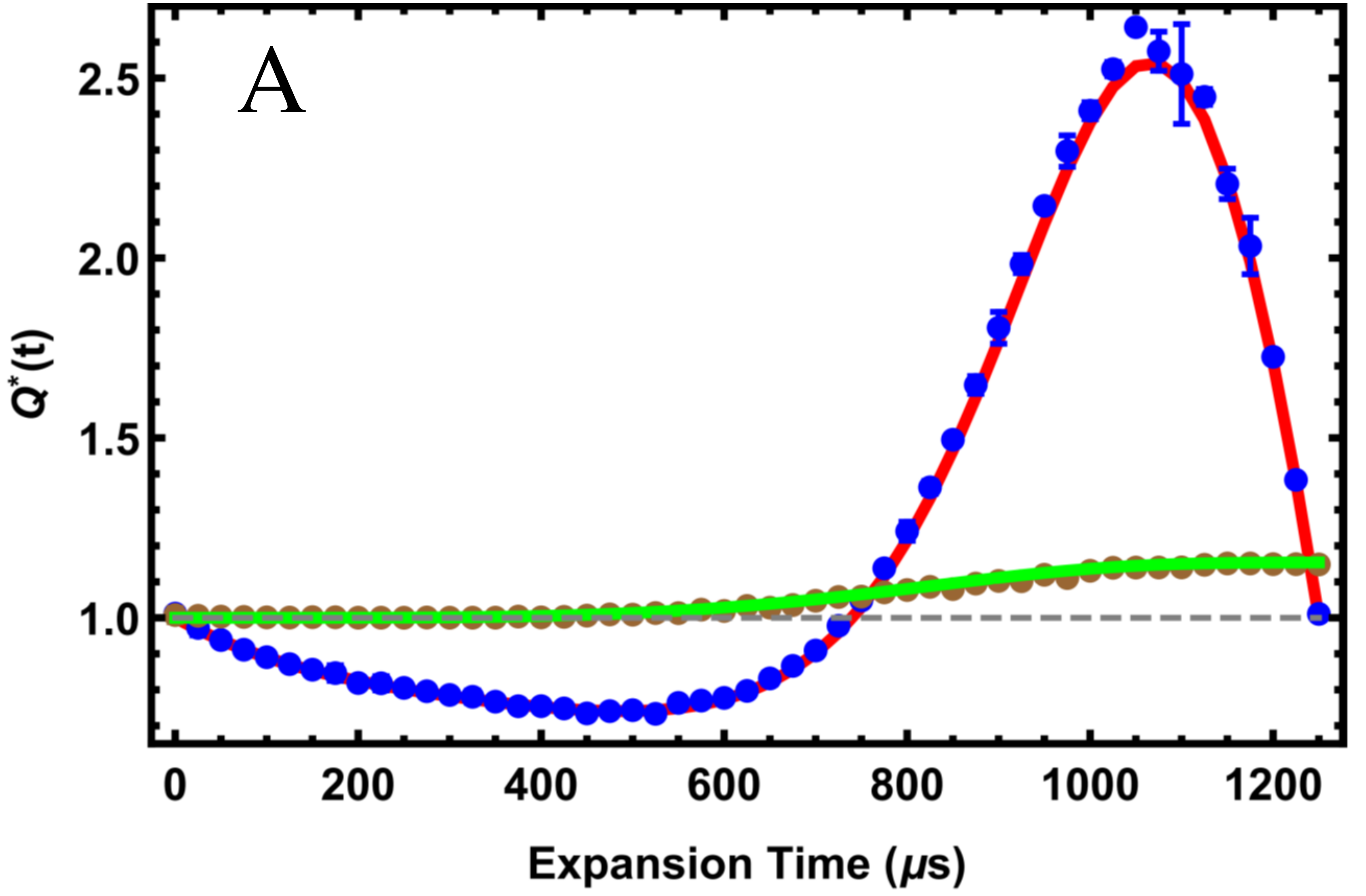}
}
\hspace{0.5cm}
\subfigure{ \label{fig1C}
\includegraphics[width=8cm]{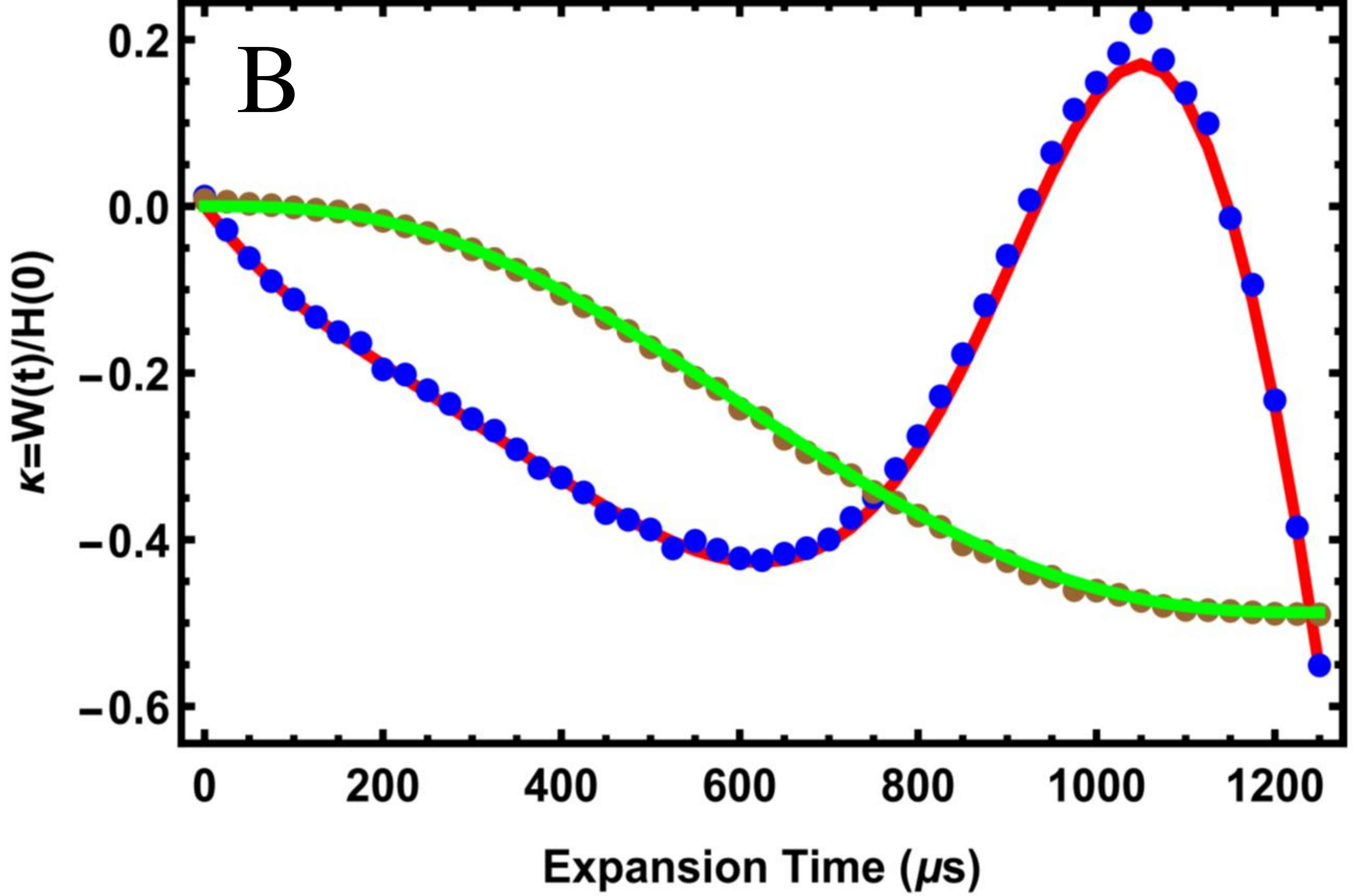}
}
\caption{{\bf Characterization of a STA for an isotropic expansion.}  (A)  Nonadiabatic factor $Q^{\ast}$ and (B) mean work . Blue  and brown dots represent measured data for LCD and reference driving, respectively,  while the red  and greens line are the corresponding theoretical predictions.}\label{QandWork}
\end{figure}

The evolution of the mean energy and mean work are  also measured in this isotropic expansion and are  shown in Fig. \ref{QandWork}. For LCD, the non-adiabatic factor $Q^*$ exhibits large deviations from unity --the adiabatic value-- evidencing the nonadiabatic character of the evolution during the STA. Nonetheless, the final value at the transferring time $\tau$ equals unity, $Q^*(\tau)=1$,  revealing a friction-free transferring process at the end of the stroke. By contrast, for the chosen reference trajectory, $Q^*$ gradually increases during the evolution and $Q^*(\tau)>1$ upon completion of the protocol. Values of $Q^*(\tau)>1$ for the reference driving indicate the presence of nonadiabatic excitations in the final state that can be associated with friction, as they are responsible for reducing the  work output with respect to the LCD, see  Fig. \ref{fig1C}.

%%%%%%%%%%%%%%%%%%%%%%%%%%%%%%
%%%%%%%%%%%%%%%%%%%%%%%%%%%%%%

\section{Shortcuts to adiabaticity  for a strongly interacting Fermi gas at high temperature}\label{sec2a}
Our implementation of STA at low temperatures relies on the existence of scale-invariance as manifested by equation (\ref{eq7}) characterizing a superfluid Fermi gas. However, the hydrodynamics can exhibit quite a different behavior in the  high-temperature regime, on which we focus next. The viscosity in this regime  modifies substantially the dynamics and thus cannot  be neglected. The cloud expansion and collective modes have been used to measure shear viscosity in the unitary Fermi gas~\cite{Cao11,Elliott14}.  To describe the dynamics in the high-temperature regime,  viscous hydrodynamics  has been used in the scaling approximation~\cite{Cao11,Elliott14,Schaefer10}.  The modified equations of motion for the scaling factors  take the form \cite{Elliott14}
\begin{equation}
\ddot{b}_j=\frac{\omega_{j,0}^2}{\Gamma^{2/3}b_j}[1+C_Q(t)]-\frac{\hbar\langle\alpha_S\rangle\sigma_{jj}}{m\langle x_j^2\rangle_0b_j}-\omega_j^2(t)b_j,\label{eq4}
\end{equation}
where the coefficient $C_Q(t)$ is the fractional increase in the volume-integrated pressure arising from viscous heating and $\langle\alpha_S\rangle$ denotes the cloud-averaged shear viscosity coefficient, $\langle \dots \rangle$ being the average over the cloud density. The coefficients $C_Q(t)$ and diagonal elements of the viscous stress tensor $\sigma_{jj}$ are specifically given by
 \begin{eqnarray}
\dot{C}_Q(t)&=&\frac{\Gamma^{2/3}}{\langle\textbf{r}\cdot\nabla U_{total}\rangle_0}\hbar\langle\alpha_S\rangle\sum_j \sigma_{jj}^2,\\
\sigma_{jj}&=&2 \left(\frac{\partial v_j}{\partial x_j} - \frac{1}{3} \nabla \cdot{\bf v} \right) = 2\frac{\dot{b}_j}{b_j}-\frac{2}{3}\frac{\dot{\Gamma}}{\Gamma} ,
\end{eqnarray}
since ${\partial v_j}/{\partial x_j} = x_j \dot{b}_j / b_j$ and $\nabla \cdot {\bf v} = \dot{\Gamma}/\Gamma$.

Note that both the viscosity heating rate coefficient $C_Q(t)$  and $\sigma_{jj}$ are zero for an isotropic expansion  with $b_x=b_y=b_z$. In this case, the equations of motion for the scaling factors  given in Eq. (\ref{eq4}) reduce to those of the superfluid unitary Fermi gas in Eq. (\ref{eq7}).  Therefore the dynamical evolution of the cloud size is then energy-independent and STA  for isotropic expansions and compressions can be efficiently implemented via LCD in this regime, with the same protocols demonstrated in the previous section. 
Nonetheless, the  time-of-flight dynamics used to probe the cloud upon completion of the STA  is modified. This is the case as the time evolution after switching off the trap is anisotropic.   The presence of shear viscosity leads then to momentum transfer from the quickly expanding direction into the slowly expanding direction. This results in  a slow decrease of the aspect ratio compared to the expansion in the superfluid regime.

Here we implement the LCD STA  to study the nonadiabatic dynamics in the high-temperature regime at unitarity. This is equivalent to a  hot superadiabatic dynamics  as those proposed for friction-free quantum thermal machines  \cite{Campo2014a,Beau16}. For simplicity, we consider an isotropic expansion stroke with the reference frequency by controlling the frequency aspect ratio, where the frequencies are designed by
\begin{eqnarray}
\omega_j(t)&=&\omega_{j,0}\bigg[1+10[b_{j}(\tau)^{-2}-1]\left(\frac{t}{\tau}\right)^3-15[b_{j}(\tau)^{-2}-1]\left(\frac{t}{\tau}\right)^4+6[b_{j}(\tau)^{-2}-1]\left(\frac{t}{\tau}\right)^5\bigg],\label{eq6}
\end{eqnarray}
where the expansion factor $b_{j}(\tau)$ is set as 1.5 and the transferring time $\tau=1.5\,ms$.

%Experimentally, we observe the evolution of the mean square cloud size by suddenly switching off the trap when we study the dynamics of the time-of-flight, which corresponds to make $\omega_j(t)$ in equation \ref{eq4} to be zero in mathematically. To reduce the unharmonic influence of optical dipole trap, we construct a crossed-dipole trap similar to our previous work\cite{STA,Efimov}. The resulting potential has a cylindrical symmetry around z axis, and the minimum trap anisotropic frequency ratio $\omega_{0r}/\omega_{0z}$ is about 9.

In this experiment, the trap depth is increased and the  aspect ratio of the trap frequencies is about 22. The system is initially prepared in a stationary state with harmonic trap frequencies $\omega_x=\omega_y=2\pi\times5581.5$ Hz and $\omega_z=2\pi\times252.7$ Hz. The harmonic trap potential $U_0$ is up to 229$\mu$K while the Fermi energy is only about 6.5$\mu$K. With this setup, the anharmonic features  of the trap are greatly suppressed. The initial energy of the Fermi gas at unitarity is $E=0.78 (0.1)\,E_F$, corresponding to a temperature $T=0.24(0.02)\,T_F$. %, where $E_F$ and $T_F$ are the Fermi energy and temperature of an ideal Fermi gas, respectively.

Subsequently, the trap frequency is lowered by decreasing the laser intensity according to Eq. (\ref{eq5}) and Eq. (\ref{eq6}), and the trap anisotropy is precisely controlled by the power ratio of the two trap beams~\cite{Deng18Sci}. Finally, after a time of evolution in the time-dependent trap, the trap beams are completely turned off and the cloud is probed via standard resonant absorption imaging techniques after a time-of-fight (TOF) expansion time $t_{TOF}$ = 500 $\mu$s. Each data point is an averaged over 5 shots taken with identical parameters. To prepare a higher temperature Fermi gas for comparison, the Fermi gas is parametrically heated up to $E=2.47\,E_F$ (corresponding to a temperature $T=0.85\,T_F$) with the same trap potential. Specifically, this is achieved by modulating the trap frequency with the resonant frequency. The time-of-flight density profile along  each direction is fitted by a Gaussian function as $A_0+A_1\exp{(-x_j^2/\sigma_j^2)}$. From this fit, we obtain the observed cloud size $\sigma_{z,obs}$ and $\sigma_{r,obs}$ that we use to determine the  $in{\text -}trap$ cloud size $\sigma_{r,in{\text -}trap}(0)$ and $\sigma_{z,in{\text -}trap}(0)$ with the hydrodynamics theory.

%The dimensionless cloud size $\widetilde{\sigma}_z$ and $\widetilde{\sigma}_r$ shown in Fig. \ref{fig3} is corrected by its initial $in{\text -}trap$ value $\widetilde{\sigma}_z=\sigma_{z,obs}/\sigma_{z,in{\text -}trap}(0),\widetilde{\sigma}_r=\sigma_{r,obs}/\sigma_{r,in{\text -}trap}(0)$.
%%%%%%%%%%%%%%%%%%%%%%%%%%%%%%
\begin{figure}[t]
\centering
\subfigure{ \label{fig3A}
\includegraphics[width=8cm]{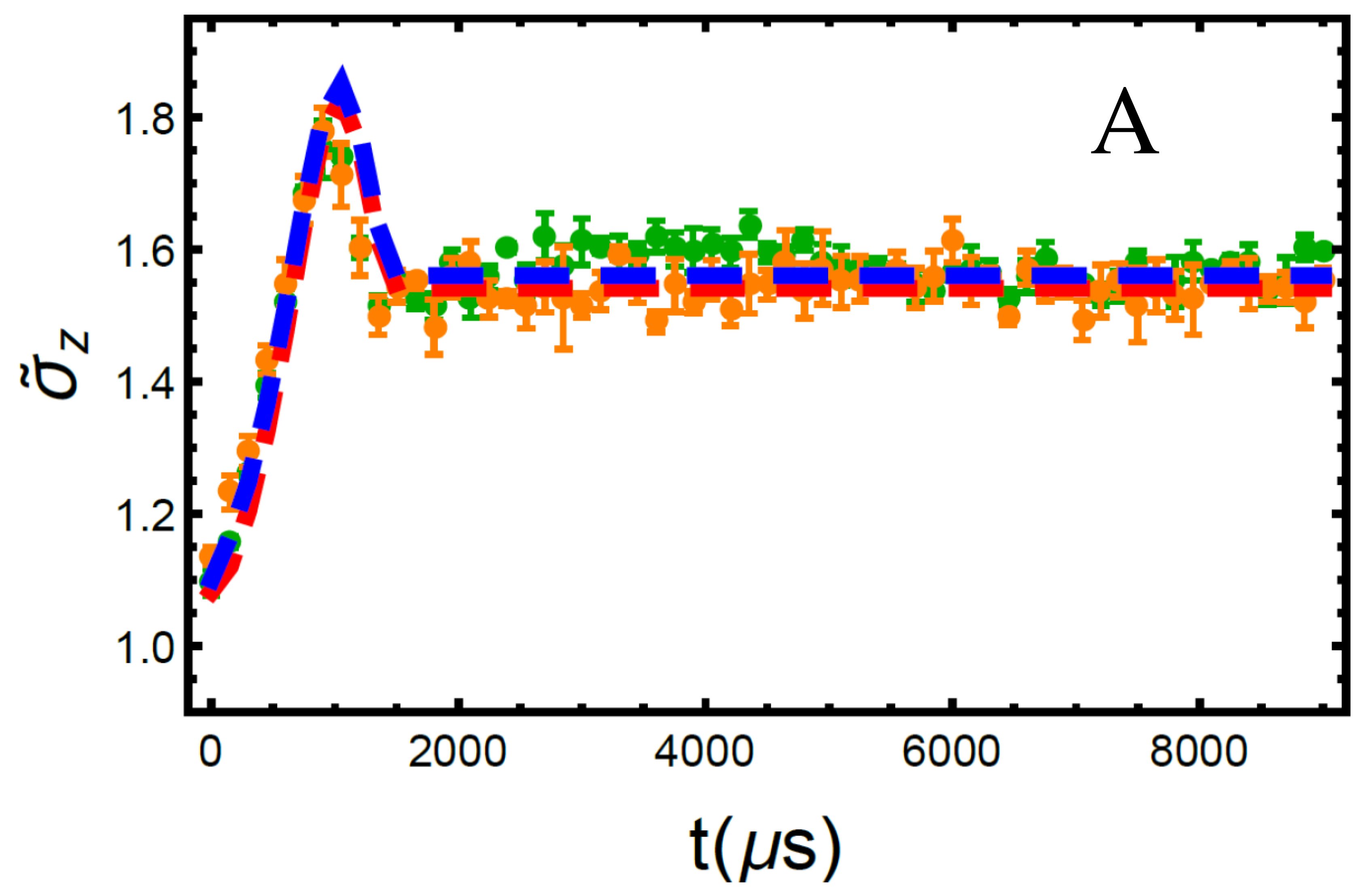}
}
\hspace{0.5cm}
\subfigure{ \label{fig3B}
\includegraphics[width=8cm]{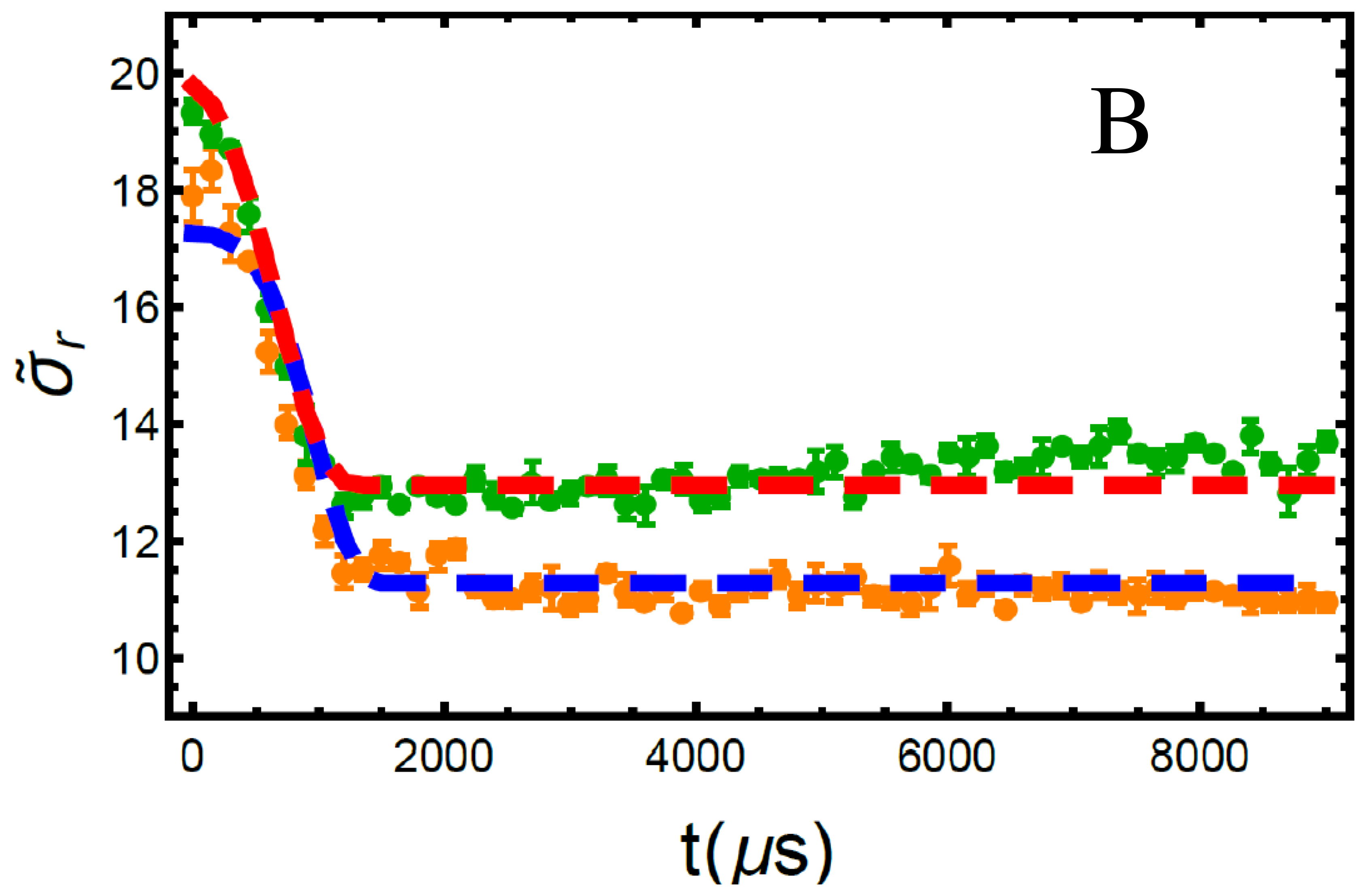}
}
\caption{{\bf Time-of-flight evolution of the cloud size following an isotropic STA}. Figure A (B) shows the dynamic behavior of the dimensionless cloud size $\widetilde{\sigma}_z(\widetilde{\sigma}_r)$ in the axial (radial) direction, where $\widetilde{\sigma}_z=\sigma_{z,obs}/\sigma_{z,in{\text -}trap}(0),\widetilde{\sigma}_r=\sigma_{r,obs}/\sigma_{r,in{\text -}trap}(0)$. The STA process starts from time $t=0$ and ends at time $t=1500\mu$s. Orange  and green dots represent  measured data with  energy $E=2.47\,E_F$  and $E=0.78\,E_F$, respectively, while the red dashed line and blue dashed line are the corresponding theoretical prediction.}\label{fig3}
\end{figure}
%%%%%%%%%%%%%%%%%%%%%%%%%%%%%%

In order to investigate the effect of shear viscosity on the dynamics at high temperature, we perform two types of experiments. We first observe the evolution of the mean square cloud size at different temperatures by suddenly switching off the trap after an isotropic STA expansion, i.e.,  implementing a TOF expansion, which corresponds to setting $\omega_j(t)$ for $(t>\tau)$ to zero in Equation (\ref{eq4}). In this case, both the viscosity heating rate coefficient $C_Q(t)$  and $\sigma_{jj}$ are zero with $b_x(t)=b_y(t)=b_z(t)$ for $t>\tau$. Isotropic STA protocol for a high-temperature Fermi gas, in principle, should be the same with the superfluid situation. However, the presence of shear viscosity will lead to momentum transfer from the radial direction into axial direction and result in the TOF dynamics quite different for different temperatures.
The  TOF expansion at the energy $E=2.47\,E_F$  and $E=0.78\,E_F$ are shown in Fig. \ref{fig3}.  After releasing the atomic cloud from the cigar-shaped trap, the shear viscosity slows the flow in the initially narrow, rapidly expanding, $x$ direction and transfers energy to the more slowly expanding $z$ direction. For a fixed time after release, the cloud aspect ratio then decreases with increasing shear viscosity. Due to the large anisotropic frequency ratio,  the expansion along the axial direction is very small and, as a result, does not exhibit significant variations for different energies, see Fig. \ref{fig3A}. However, the gas experiences fast expansion along the radial direction, reaching a size about 20 times bigger than the initial one.  This illustrates clearly the effect of increasing the shear viscosity, see Fig. \ref{fig3B}.  The small residual excitation following  the STA is because  the engineered  frequency in the experiment differs slightly from the designed ideal trajectory.

In a second kind of experiment in the high-temperature regime, we investigate the influence of the shear viscosity on the anisotropic expansion.  For a cylindrical symmetric dipole trap, the frequencies $\omega_x$ and $\omega_y$ should always be the same, meaning that the scale factors fulfill $b_x(t)=b_y(t)$.  Referring to Eq. (\ref{eq4}) to implement a STA in a high temperature of the unitary Fermi gas, the frequencies should  satisfy
\begin{eqnarray}
\Omega_j^2(t)&=&\frac{\omega_{j,0}^2}{\Gamma^{2/3}b_j^2}\left[1+C_Q(t)\right]-\frac{\hbar\langle\alpha_S\rangle\sigma_{jj}}{m\langle x_j^2\rangle_0b_j^2}-\frac{\ddot{b}_j}{b_j}.\label{eq8}
\end{eqnarray}
 Here the trap-averaged shear viscosity and the viscous heating coefficient $C_Q(t)$  need to be determined to design the trap frequencies and  aspect ratio. Although they have been precisely measured in equilibrium, the dynamics of the trap-averaged shear viscosity is very complex. As a result, we implement a STA by LCD that is
guaranteed to work for a unitary Fermi gas with no viscosity, using Eqs. (\ref{eq7}) and (\ref{LCDfreqanis}),  and study the deviations that arise due to the viscous
hydrodynamics. 
To this end, we compare the dynamics in both isotropic and anisotropic STA protocols,  for which the trap frequencies  are chosen as follows
\begin{eqnarray}
{\text {Isotropic STA}}~~\omega_z(0)&=&2\pi\times252.7{~\text {Hz}}~\longrightarrow\omega_z(\tau)=2\pi\times112.3{~\text {Hz}},\nonumber\\
~~\omega_x(0)&=&2\pi\times5581.5{~\text {Hz}}\longrightarrow\omega_x(\tau)=2\pi\times2480.7{~\text {Hz}},\nonumber\\
{\text {Anisotropic STA}}~~\omega_z(0)&=&2\pi\times252.7{~\text {Hz}}~\longrightarrow\omega_z(\tau)=2\pi\times208.8{~\text {Hz}},\nonumber\\
~~\omega_x(0)&=&2\pi\times5581.5{~\text {Hz}}\longrightarrow\omega_x(\tau)=2\pi\times2480.7{~\text {Hz}}.\nonumber
\end{eqnarray}
%
%%%%%%%%%%%%%%%%%%%%%%%%%%%%%%
\begin{figure}[t]
\centering
\includegraphics[width=16cm]{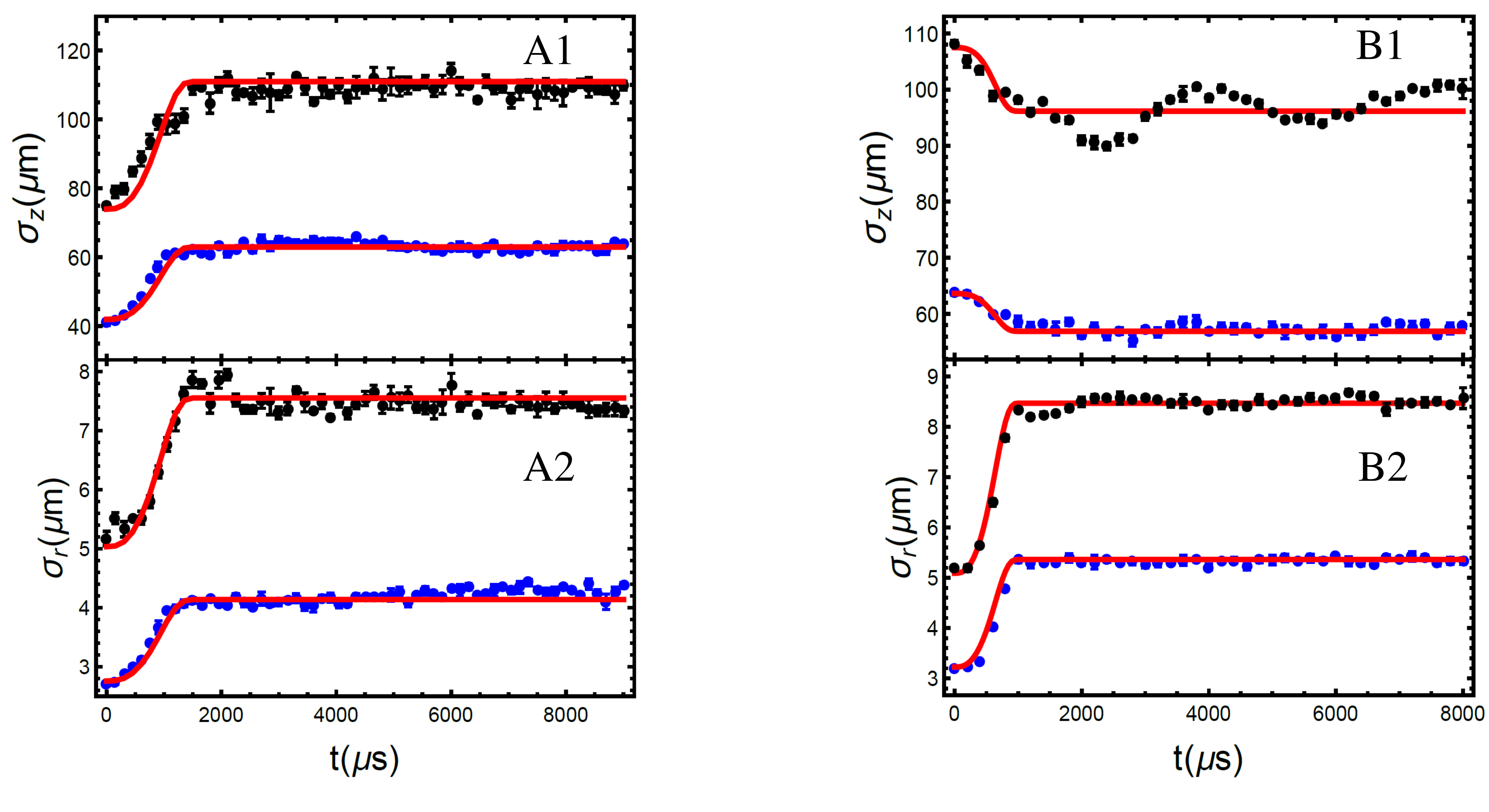}
\caption{{\bf Dynamical evolution of a unitary Fermi gas at different temperatures.} Figure A1 (A2) shows the evolution of the cloud size along the axial (radial) direction for an isotropic expansion while Figure B1 (B2) indicates the dynamic behavior of the cloud size along the axial (radial) direction for an anisotropic expansion. Black (blue) dots are measured at high (low) temperature corresponding to an initial energy $E=2.47(0.78)\,E_F$ and the solid lines are corresponding theoretical predictions without considering the viscosity.}\label{fig4}
\end{figure}
%%%%%%%%%%%%%%%%%%%%%%%%%%%%%%
The aspect ratio of the target stationary state for an anisotropic expansion is 11.9. For comparison, the STA trajectories are implemented with the energies of $E=2.47\,E_F$ and $E=0.78\,E_F$, respectively, see Fig. \ref{fig4}.
For the isotropic expansion, the dynamics along different directions shares the same behavior, shown in Fig. \ref{fig4} A1 and A2. The viscosity rarely affects the dynamic evolution even at a quite high temperature with energy values up to $2.47\,E_F$. By contrast, the anisotropic expansion dynamics, where $b_x=b_y\neq b_z$, shows different behavior with increasing viscosity for different energy values. The STA for the anisotropic expansion works well at low temperatures. In the strongly coupled regime, the cloud size in the axial direction behaves as in a compression stroke, since the frequency in the radial direction decreases faster and the energy would ``flow" into the radial direction. The experimental results are consistent with the theoretical calculation by Eq. (\ref{eq7}). However the dynamical behavior of the axial direction exhibits an excitation at high temperature while the radial behavior is still consistent with the theoretical prediction. The large deviation between $b_x$ and $b_z$ would result in a constant increase of the viscous heating coefficient $C_Q(t)$. When the viscosity coefficient $\langle\alpha_S\rangle$ is large and  $\frac{\hbar\langle\alpha_S\rangle\sigma_{jj}}{m\langle x_j^2\rangle_0b_j^2}$ becomes comparable to the square of the frequency, the STA trajectory should be corrected according to Eq. (\ref{eq8}). Neglecting the contribution of viscosity, the expansion stroke does not satisfy the boundary conditions and thus exhibits some excitation after the transferring time. Since the  frequency aspect ratio is very large, the contribution of the viscosity in the radial direction is  smaller than the square of the frequency. We could hardly see the deviation of the expansion behavior away from its theoretical calculation in the radial direction which is shown in Fig. \ref{fig4} B2.
%%%%%%%%%%%%%%%%%%%%%%%%%%%%%%
\begin{figure}[t]
\centering
\subfigure{ \label{fig5A}
\includegraphics[width=8cm]{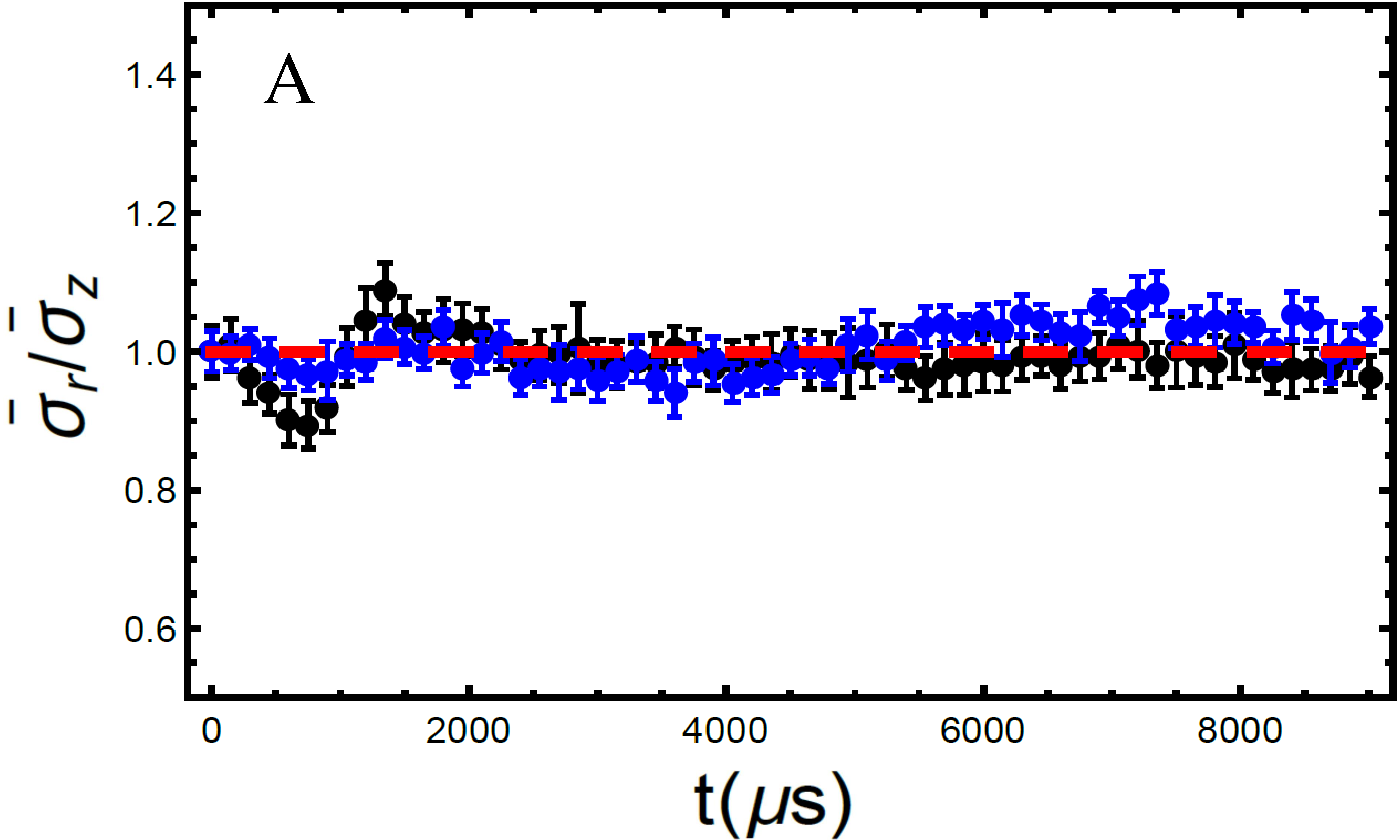}
}
\hspace{0.5cm}
\subfigure{ \label{fig5B}
\includegraphics[width=8cm]{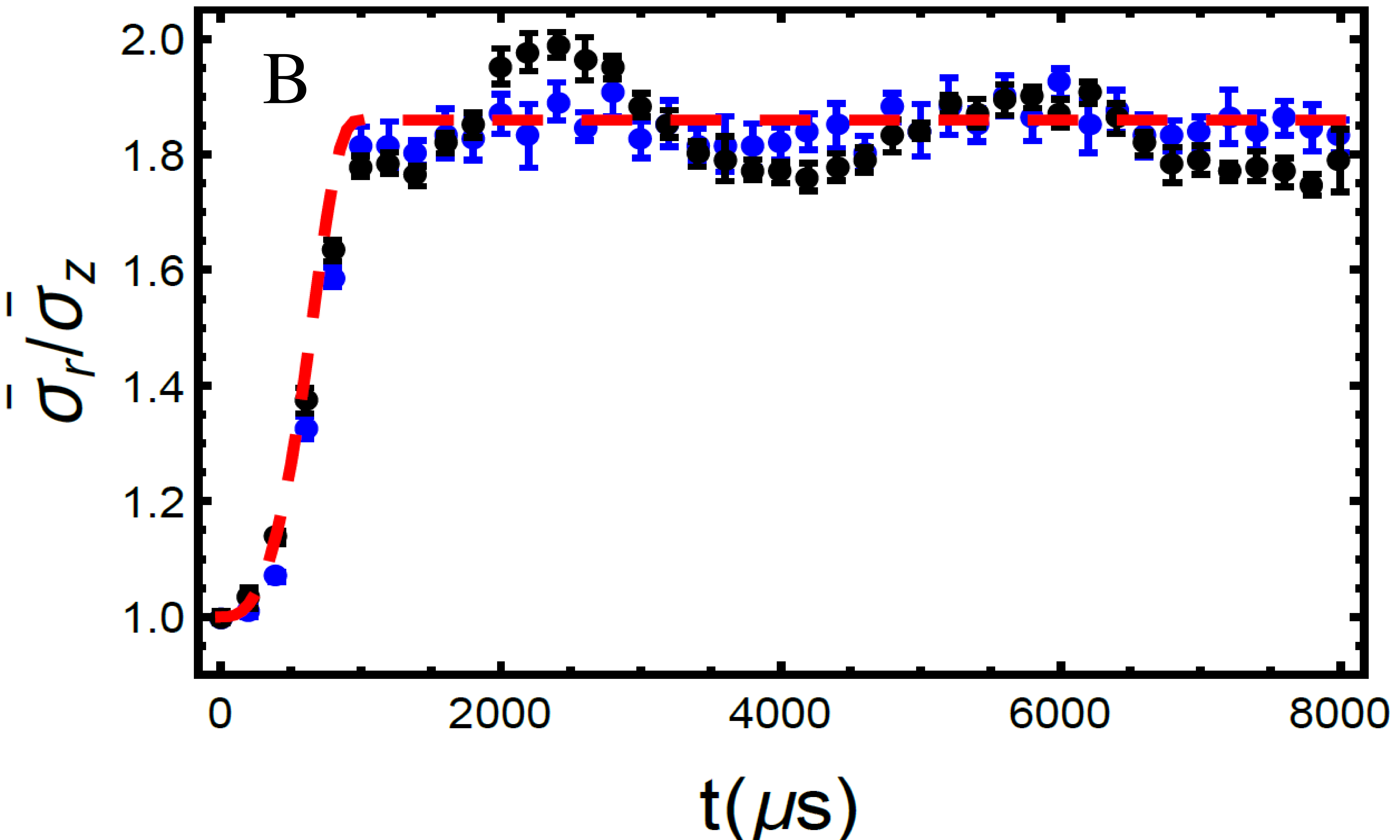}
}
\caption{{\bf Evolution of the dimensionless ratio $\bar{\sigma}_r/\bar{\sigma}_z$ at different temperatures.} (A) Isotropic and (B) anisotropic shortcut to an adiabatic expansion. Here, $\bar{\sigma}_r=\sigma_r(t)/\sigma_{r}(0)$ and $\bar{\sigma}_z=\sigma_z(t)/\sigma_{z}(0)$. Blue  dots are measured at low temperature with initial energy $E=0.78 \,E_F$ while black dots correspond to the high-temperature viscous regime with initial energy $E=2.47\,E_F$. Dashed lines denote the corresponding theoretical predictions without considering the viscosity.}\label{fig5}
\end{figure}
%%%%%%%%%%%%%%%%%%%%%%%%%%%%%%

To further compare the dynamic of the atomic cloud for the isotropic and anisotropic expansion, the dimensionless cloud size $\bar{\sigma}_j=\sigma_j(t)/\sigma_j(0)$ is shown in Figure \ref{fig5}.  The ratio of $\bar{\sigma}_r/\bar{\sigma}_z$ is very close to one and the system remains at thermal equilibrium when the STA driving is completed. For different energies at different times, the ratio $\bar{\sigma}_r/\bar{\sigma}_z$ keeps a constant value closed to unity as shown in Fig. \ref{fig5A}. The slightly deviation of the black dots during the nonadiabatic transferring time is due to the large viscosity. The anisotropic expansion shown in Fig. \ref{fig5B} is largely dependent on the energy. When the energy is low and the viscosity can be neglected, the ratio $\bar{\sigma}_r/\bar{\sigma}_z$  keeps  its  constant aspect ratio $b_{x}(\tau)/b_{x}(0)=1.86$. However, $\bar{\sigma}_r/\bar{\sigma}_z$ oscillates for high energy due to the presence of the viscosity. Further, contrary to the superfluid case, residual excitations of the breathing mode are damped as a function of time due to the viscous hydrodynamics.

\section{Conclusions}\label{sec3}

In conclusion, we have studied the control of the  nonadiabatic expansion dynamics of an interacting Fermi gas   in both the noninteracting  and unitary regimes.
To this end, we have engineered shortcuts to adiabaticity by counterdiabatic driving exploiting scale-invariance as an emergent dynamical symmetry in these two limits.
By doing so, the cloud size  follows a prescribed adiabatic trajectory without the requirement of slow driving that can be used to implement a superadiabatic transition between two different stationary quantum states. Superadiabatic expansions can be applied in a variety of scenarios, and can be used as a dynamical microscope to probe the  state of the atomic cloud \cite{Campo2011,Papoular15} as well as to implement friction-free superadiabatic strokes in quantum thermodynamics \cite{Deng13,Campo2014a,Beau16,Funo17}.

For the 3D anisotropic ideal Fermi gas, we have implemented shortcuts to adiabaticity  via an isotropic nonadiabatic expansion. To this end we have engineered common  scaling factor describing the expansion of the atomic cloud along all different axes as a function of time. This is possible thanks to the individual control of the trap frequencies as well as their aspect ratio that allow for the implementation of a shortcut to adiabaticity even in the resonant regime, using the technique proposed in \cite{Deng18Sci}.

We have also investigated shortcuts to adiabaticity at high temperature for a  unitary Fermi gas in a time-dependent anisotropic Fermi gas. The time-of-flight dynamics is changed as the increasing shear viscosity transfers the momentum from the quickly expanding direction into the slowly expanding direction. By comparing the dynamical evolution along a shortcut to adiabaticity for   isotropic  and anisotropic expansions, we have demonstrated the impact of the shear viscosity on  the nonadiabatic dynamics and its effect on the residual excitation of the breathing modes of the cloud.

\section*{Acknowledgements}

This research is supported by the National Key Research and Development Program of China (grant no.2017YFA0304201), National Natural Science Foundation of China (NSFC) (grant nos. 11734008, 11374101, 91536112, and 11621404), Program of Shanghai Subject Chief Scientist (17XD1401500), the Shanghai Committee of Science and Technology (17JC1400500), UMass Boston (project P20150000029279) and the John Templeton Foundation. AdC acknowledges 
partial support from Institut Henri Poincar\'e  and  CNRS via the thematic trimester at the Centre  \'Emile Borel  entitled
``Measurement and control of quantum systems: theory and experiments'' in Spring 2018.

\bibliographystyle{iopart-num}

\bibliography{STAFermi}

\end{document}